\documentclass{nature}
\usepackage{graphicx}
\usepackage{amsmath}
\usepackage{amssymb}
\usepackage[usenames]{color}

\title{Extreme cooperative swelling in topologically disordered fibre entanglements}

\author{Alistair R. Overy,$^{1,2}$ Raj Pandya,$^1$ Phillip M. Maffettone,$^1$ Philip A. Chater,$^2$ Arkadiy Simonov$^1$ \& Andrew L. Goodwin$^{1,\ast}$}

\begin{document}

\maketitle

\begin{affiliations}
 \item Department of Chemistry, University of Oxford, Inorganic Chemistry Laboratory, South Parks Road, OX1 3QR, Oxford, UK
 \item Diamond Light Source, Chilton, Oxfordshire, OX11 0DE, UK
\end{affiliations}

\begin{abstract}
Entangled states are ubiquitous amongst fibrous materials, whether naturally occurring (keratin, collagen, DNA) or synthetic (nanotube assemblies, elastane). A key mechanical characteristic of these systems is their ability to reorganise in response to external stimuli, as implicated in \emph{e.g.}\ hydration-induced swelling of keratin fibrils in human skin. During swelling, the curvature of individual fibres changes to give a cooperative and reversible structural reorganisation that opens up a pore network. The phenomenon is known to be highly dependent on topology, even if the nature of this dependence is not well understood: certain ordered entanglements (`weavings') can swell to many times their original volume while others are entirely incapable of swelling at all. Given this sensitivity to topology, it is puzzling how the disordered entanglements of many real materials manage to support cooperative dilation mechanisms. Here we use a combination of geometric and lattice-dynamical modelling to study the effect of disorder on swelling behaviour. The model system we devise spans a continuum of disordered topologies and is bounded by ordered states whose swelling behaviour is already known to be either vanishingly small or extreme. We find that while topological disorder often quenches swelling behaviour, certain disordered states possess a surprisingly large swelling capacity. Crucially, we show that the extreme swelling response previously observed only for certain specific weavings can be matched---and even superseded---by that of disordered entanglements. Our results establish a counterintuitive link between topological disorder and mechanical flexibility that has implications not only for polymer science but also for our broader understanding of collective phenomena in disordered systems.
\end{abstract}

The packing of one-dimensional objects in three-dimensional space is a non-trivial geometric problem of relevance to the phase behaviour of synthetic polymers,\cite{Baughman_1973} carbohydrate fibres,\cite{Kong_2012} carbon nanotubes,\cite{Hall_2008,Hu_2010} biomaterials,\cite{Brody_1960,Stokes_1991,Norlen_2004} liquid crystals,\cite{Spicer_2001,Mol_2005,Zeng_2011} lipid assemblies,\cite{Saludjian_1980} superconductors,\cite{Abrikosov_1957,Auslaender_2009} skyrmion magnets,\cite{Muhlbauer_2009} and hybrid inorganic--organic frameworks.\cite{Rosi_2005} The simplest possible arrangements are the rod packings,\cite{OKeeffe_1977,OKeeffe_2001} where packing symmetry relates the position and orientation of each rod to those of its neighbours [Fig.~1(a)--(c)]. Rods have no curvature, but fibres may; curvature allows for denser assemblies and its introduction into the rod packings transforms them into periodic `weavings' with the same underlying topology yet different packing density.\cite{Evans_2011,Evans_2013,Evans_2015} This transformation reflects the mechanism by which physical processes that affect curvature---such as hydration\cite{Norlen_2004} or thermal treatment\cite{Baughman_1973}---can in turn control the volume of fibre assemblies without requiring changes in topology. The ability of a particular weaving to swell as curvature varies is quantified by the ratio $\tau=V_{\rm max}/V_{\rm min}$ between the unit cell volumes of the extended (`max') and dense (`min') geometries. The magnitude of $\tau$ is a function of weaving topology.\cite{Evans_2011} For the ubiquitous hexagonal fibre packing, $\tau=1$; it cannot swell. This is true also for the simple cubic packing $\Pi^\ast$ (itself related to the crystal structure of elemental W).\cite{OKeeffe_1977} By contrast, the $\Sigma^\ast$ packing---which describes the arrangement of channels in gyroid membrane structures---admits an extreme swelling ratio $\tau=5.429$ that is comparable to the effect observed experimentally in key biopolymers [Fig.~1(c),(d)].\cite{Norlen_2004} In other words, whereas the process of varying fibre curvature can result in a five-fold volume increase for weavings with the $\Sigma^\ast$ topology, the same mechanism has no effect on other fibre arrangements, such as those related to the hexagonal or $\Pi^\ast$ rod packings.


An inevitable corollary of this specificity is that the swelling ability of any fibre assembly must be sensitive to the presence of topological disorder. This argument is entirely consistent with intuition: one expects the cooperative unwinding mechanism responsible for dilation to be locked as disorder introduces knots---perhaps even at extremely low concentrations.\cite{Evans_2011} Optical microscopy clearly reveals the presence of varying degrees of topological disorder within a variety of synthetic and natural fibre packings for which swelling is considered important [Fig.~1(e)]. By convention, such systems (we use the term `entanglements' hereafter) are generally modelled in terms of two independent components: an ordered weaving whose topology reflects structural motifs present locally within the fibre assembly, and a separate random-coil fraction.\cite{Segerman_1966} Hence there is a fundamental disconnect between our understanding that mechanical response is so strongly topology dependent, on the one hand, and the approximation of real systems in terms of a non-interacting two-phase model, on the other hand. Moreover the recent discovery of reversible dilatancy in `random' coils\cite{Rodney_2016} together with the longer-established identification of negative Poisson ratios in amorphous foams\cite{Alderson_2007} collectively illustrate the propensity for topologically disordered networks to exhibit mechanical responses that are difficult to anticipate based on the consideration of ordered states alone. So the relationship between topological disorder in fibre entanglements and cooperative mechanical responses such as swelling is an open, interesting, and important question.
 

Here we study this relationship using a combination of geometric and lattice-dynamical approaches. A key challenge is to devise a methodology that allows topological disorder to be controlled systematically such that disorder might be correlated directly with mechanical response. We identify a common topological parent of both $\Pi^\ast$ and $\Sigma^\ast$ weavings which allows us to generate intermediate entanglements of essentially arbitrary (but quantifiable) degrees of topological disorder. Moreover, because the swelling capabilities of the bounding topologies are so very different---\emph{i.e.}\ vanishingly small in the case of $\Pi^\ast$ and extreme in the case of $\Sigma^\ast$---we can anticipate that $\tau$ must vary with degree of topological disorder. We use lattice-dynamical calculations to measure $\tau$ as a function of entanglement topology, benchmarking our results against the exact values of the limiting cases $\Pi^\ast$ and $\Sigma^\ast$.\cite{Evans_2011} We show that the swelling capacity of the $\Sigma^\ast$ weaving is indeed sensitive to disorder, but that certain disordered topologies allow recovery of extreme swelling behaviour. Indeed our key result is the discovery of disordered entanglements that are capable of swelling to even greater extents than the $\Sigma^\ast$ weaving itself (\emph{i.e.}\ $\tau\gtrsim6$). Hence we show definitively that topological order is not prerequisite for the large swelling behaviour observed in some fibrous materials, with disordered entanglements capable of supporting the collective rearrangements required for extreme and counterintuitive mechanical response.


We begin by describing our methodology for generating microscopic models of entangled states. These models will later form the input for lattice-dynamical calculations to determine swelling behaviour. Our model generation approach consists of two steps: first we obtain a topological representation of each entanglement with a quantifiable degree of disorder, and second we embed this topological state as an atomistic configuration, the geometry of which is then optimised using a set of simple pairwise interaction potentials.

In Fig.~2(a) we illustrate a single unit cell (space group symmetry $Ia\bar3d$) of the parent network from which our various entanglements are derived. The nodes of this network occupy the $48g$ Wyckoff sites and are connected by two symmetry-inequivalent types of linker [shown in red and teal in Fig.~2(a)]. Four linkers meet at each node: two of each type. Entanglements are obtained by partitioning the linkers into equal sets such that each set contains exactly two linkers meeting at every node (see SI for further details). Consequently a given linker set describes a family of self-avoiding one-dimensional paths embedded on the network, which is dense with respect to coverage of the nodes; we associate with each set its constituent fraction $x$ of teal linkers. The trivial partitioning is the single possibility that completely respects linker type, giving entanglements with the $\Pi^\ast$ ($x=0$; \emph{i.e.}\ all red linkers) and $\Sigma^\ast$ ($x=1$; \emph{i.e.}\ all teal linkers) topologies [Fig.~2(b),(c)]. Entanglements corresponding to intermediate values of $x$ represent approximants to disordered states, consistent by design with the periodic boundary conditions imposed by the underlying network unit cell [Fig.~2(d)]. The use of larger supercells gives increasingly realistic approximants but at considerable computational cost. The topologies generated using this approach are related both to the `spaghetti states' of Coulomb phases such as the spin ices\cite{Henley_2010,Moessner_2006} and also to the procrystalline networks of Ref.~\citenum{Overy_2016} as realised in \emph{e.g.}\ transition-metal oxyfluorides.\cite{Yang_2011,Camp_2012}

Having generated an entanglement for a given value of $x$, the corresponding linker set is decorated by atoms to give a crude atomistic representation of a fibre structure with the same topology (see SI). Finally, the atom positions are relaxed subject to a combination of harmonic bond potentials and electrostatic repulsion between fibres. The harmonic potential acts to maintain fibre connectivity while the electrostatic term maximises fibre separation; both terms---full details of which are given as SI---are implemented so as to respect the topology imprinted during set partitioning. An illustration of the result of this process is given in Fig.~2(e), with further representative relaxed configurations shown in the SI.



We proceeded to use lattice-dynamical calculations to identify the swelling behaviour for our various entanglements, our eventual goal having been to characterise the function $\tau(x)$, $0\leq x\leq1$. Before presenting our results we describe first the basic approach we have taken together with our benchmarking checks for the (well understood) $\Pi^\ast$ and $\Sigma^\ast$ topologies. For a given value of $x$ we generated a range of topologically-distinct entanglements as described above. Then, for each entanglement, we used the General Utility Lattice Program (GULP)\cite{Gale_1997} to perform a structural relaxation, which was in turn repeated for a variety of different nominal cell lengths $d$. The motivation for varying cell length was to access different states of the same topology with different degrees of swelling. Each of these relaxed configurations can be compared to one another as entanglements of unit-diameter fibres after rescaling by the closest inter-fibre separation $r$. This gives a final simulation cell of length $L=d/r$. By repeating our calculations while incrementally varying $d$, we can determine what particular range of $L$ values is accessible to a given entanglement without deforming the fibres themselves (\emph{i.e.} changes in curvature are allowed, but stretching is not). For each of these allowed $L$-states of a given entanglement, fibres are in direct contact; at the minimum value $L=L_{\rm{min}}$ the fibres have their greatest curvature and the entanglement is most densely packed, while at the maximum value $L=L_{\rm{max}}$ the system is at its most swollen with an open pore network (\emph{i.e.}\ low fibre curvature). The swelling ratio is simply a measure of the relative volumes of the dense and open states:
\begin{equation}
\tau=\left(\frac{L_{\rm{max}}}{L_{\rm{min}}}\right)^3.
\end{equation}

For the $\Pi^\ast$ weaving at $x=0$ we can access only a small range of $L$-states with $L\simeq4$ [Fig.~3(a)]. Hence $L_{\rm{min}}\simeq L_{\rm{max}}$ and $\tau(0)=1.04\simeq1$, as expected. By contrast, for the $\Sigma^\ast$ weaving at $x=1$ our GULP relaxations give $L$-states with $4.20\leq L\leq6.67$, and hence our approach estimates $\tau(1)=4.0$ [Fig.~3(b)]. An illustration of the GULP configurations across this range of $L$-states is given in Fig.~3(e). The progression from $L_{\rm min}$ to $L_{\rm max}$ clearly identifies the same cooperative swelling mechanism already established in Ref.~\citenum{Evans_2011}. For context, we note that the swelling ratio determined using our new approach is about as close to the exact result ($\tau=5.429\ldots$) as that obtained in the earlier simulations of Ref.~\citenum{Evans_2014} ($\tau=3.9$). The level of agreement we find here corresponds to a \emph{ca} 10\% uncertainty in the value of $L$, which we will come to show is sufficient for qualitative comparison of the swelling capabilities of different entanglements. In each of these calculations---as in representative examples of the disordered entanglements described below---we verified reversibility through repeated cycling of $d$ trajectories (see SI). Moreover, perturbations in the numerical parameters employed in our GULP simulations (\emph{e.g.}, harmonic potential strength, effective fibre charge density) had no meaningful effect on the critical ratio $L_{\rm max}/L_{\rm min}$, and hence on $\tau$ (see SI).

The motivation for developing this approach is that it is sufficiently general to allow estimation of $\tau(x)$ for entanglements corresponding to intermediate values $0<x<1$. A survey of our results for these intermediate disordered states is given in Fig.~4(a), which reveals a highly non-trivial relationship between topological disorder and capacity for swelling. There appear to be two competing trends at play. On the one hand, there is the anticipated continuum of states between $\Sigma^\ast$ and $\Pi^\ast$ for which disorder quickly favours the limit $\tau\rightarrow1$. So, for example, low levels of topological disorder within the $\Sigma^\ast$ topology appear to quench swelling in most cases. On the other hand, there is a broad distribution of a small fraction of entanglements for which surprisingly large values of $\tau$ are realised. Taken together, these two effects mean that at intermediate values of $x$ some entanglements show very small swelling effects while others show very large effects; we illustrate this point in Fig.~3(c),(d) by plotting the allowed $L$-states for two entanglements at $x=0.5$ with contrasting swelling abilities. Importantly, our results identify the existence of disordered entanglements with swelling ratios even larger than that of the ordered $\Sigma^\ast$ topology. In Fig.~3(f) we show the swelling mechanism for one such entanglement. The mechanism is clearly strongly cooperative, even if its nature is very different to that for the $\Sigma^\ast$ weaving (by way of example).

Just as the relationship between topology and swelling capability remains poorly understood even for ordered weavings,\cite{Evans_2011,Evans_2015} so too is there no obvious rationale for the variance in $\tau$ we observe for different disordered fibre entanglements. Our key point, of course, is simply that large values of $\tau$ are accessible to disordered states. Nevertheless we do make some further observations regarding our results. First, the variable $x$ does not by itself guarantee disorder for intermediate values since ordered configurations with $x=0.5$ are realisable in principle (even if statistically improbable). Nevertheless the configurations we observe to give extreme swelling responses are \emph{not} ordered, as evidenced by their trivial $P1$ symmetry. And, second, our approach for generating disordered configurations remains based on a specific geometric model and necessarily allows us to explore only a small and finite subset of the family of allowed disordered entanglements. Consequently we have established only a lower bound on the maximum swelling behaviour of this broader family. In other words, disordered entanglements of suitable topologies may in fact show superior swelling responses to ordered weavings. In terms of the implications for engineering materials with dilatant properties, our results suggest that control over the extent and type of topological disorder may be as an effective a design strategy as one of targeting specific ordered motifs.

The inverse progression---from open to dense packings---bears a conceptual similarity to the `tension' mechanism for negative thermal expansion (NTE),\cite{Mary_1996,Barron_1989} whereby a material contracts on heating. In this picture, the low-energy vibrational modes of a NTE material involve flexing of initially-linear bonding arrangements such that the system folds in on itself as the magnitude of vibration increases at high temperatures. Consequently, whether or not a material exhibits NTE largely relies on the existence of a collective mechanism of bond flexing that allows this concerted densification; this is the case, for example, in ZrW$_2$O$_8$ and Zn(CN)$_2$.\cite{Mary_1996,Goodwin_2005} As for swelling, the existence of such a mechanism is known to be a non-trivial function of network topology.\cite{Dove_2007} It is for precisely this reason that the question as to how amorphous materials (\emph{e.g.}\ silica\cite{Dove_1997} and some liquids\cite{Debenedetti_2003}) show NTE has remained one of the most important and challenging problems in the field. Our geometric model for swelling behaviour is extremely similar in that it effectively links fibre curvature with changes in cell volume. Drawing analogies to NTE, the dense packing at $L_{\rm min}$ would reflect the high-temperature state (bent linkages, low volume) and the open packing at $L_{\rm max}$ the low-temperature state (linear linkages, high volume) [Fig.~4(b)]. So our finding that disordered fibre arrangements can reduce their volume by increasing fibre curvature also establishes a heuristic explanation as to why NTE might be observed in ostensibly unrelated families of disordered materials. The extension to negative Poisson's ratios\cite{Lakes_1987,Greaves_2011}---as known for some amorphous foams\cite{Alderson_2007}---may also follow similar arguments. Whether this mapping proves robust or not, what emerges in all cases is the obvious importance of collective degrees of freedom in driving counterintuitive physical properties of topologically disordered systems.







\bibliography{arobib}

\begin{addendum}
 \item The authors gratefully acknowledge funding from the E.R.C. (Grant 279705), the Leverhulme Trust (Grant RPG-2015-292), the Swiss National Science Foundation to A.S, the Marshall Foundation to P.M.M., and from the Diamond Light Source to A.R.O. A.L.G. is grateful to M. Evans (TU Berlin) and M. Tucker (SNS, ORNL) for valuable discussions.
 \item[Competing Interests] The authors declare that they have no competing financial interests.
 \item[Correspondence] Correspondence and requests for materials should be addressed to A.L.G.~(email: andrew.goodwin@chem.ox.ac.uk).
\end{addendum}

\clearpage
\begin{center}
\includegraphics{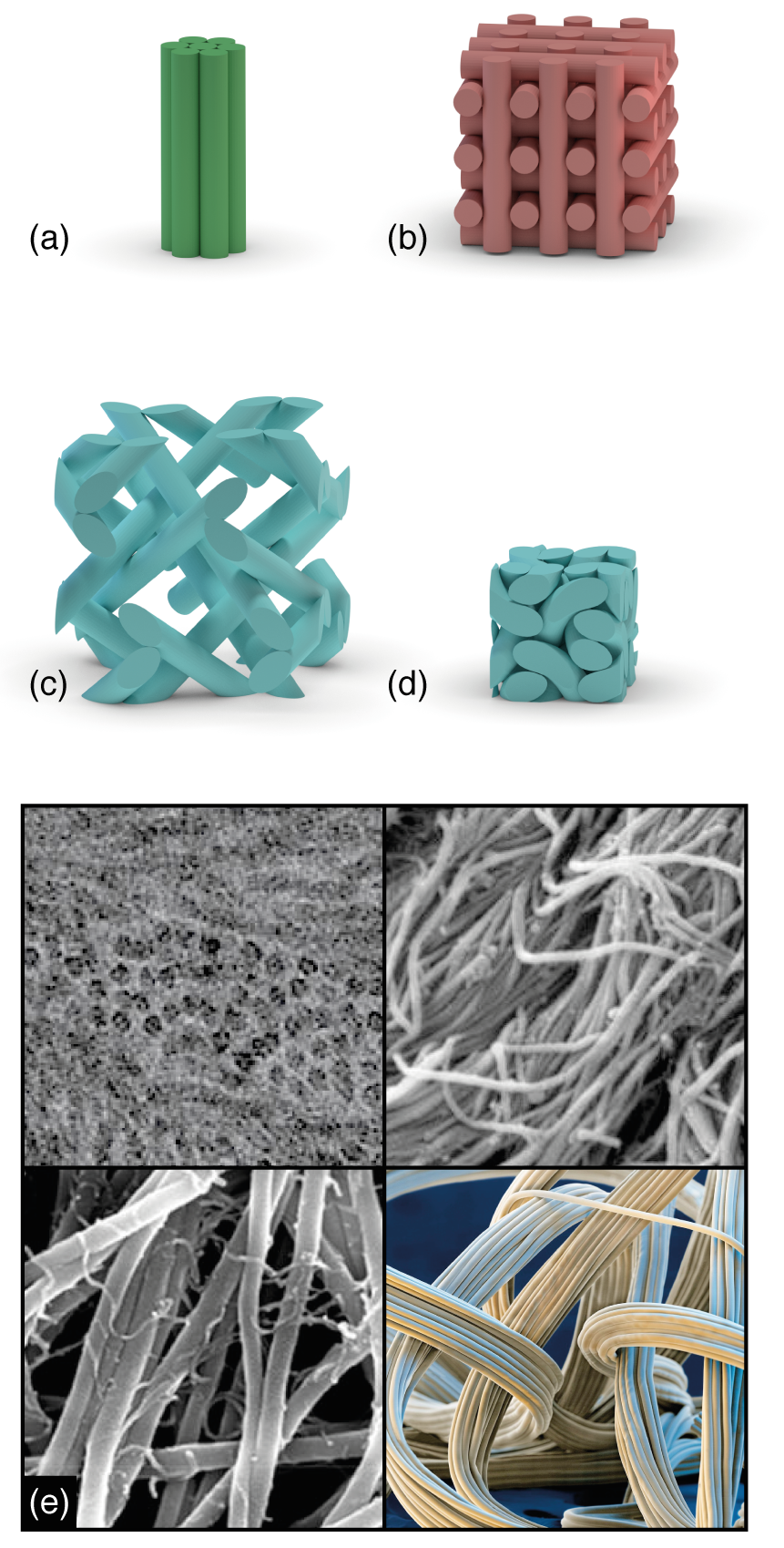}
\end{center}
{\bf Figure 1. Rod packings and entanglements.} Representations of the (a) hexagonal, (b), $\Pi^\ast$, and (c) $\Sigma^\ast$ rod packings. If curvature is allowed then the $\Sigma^\ast$ topology describes a fibre weaving with variable pore volume. A representation of this weaving in its densest state (minimal pore volume) is shown in (d). The interconversion between open (c) and closed (d) states is topologically protected and involves only a cooperative change in curvature of the individual fibres. Panel (e) gives representative microscopy images for a variety of naturally-occurring and synthetic disordered fibre entanglements: (clockwise from top-left) keratin,\cite{Norlen_2004} collagen,\cite{LarangeiradeAlmedia_2013} elastane,\cite{Spandex} and carbon nanotubes.\cite{FerreiraSantos_2015}

\clearpage
\begin{center}
\includegraphics{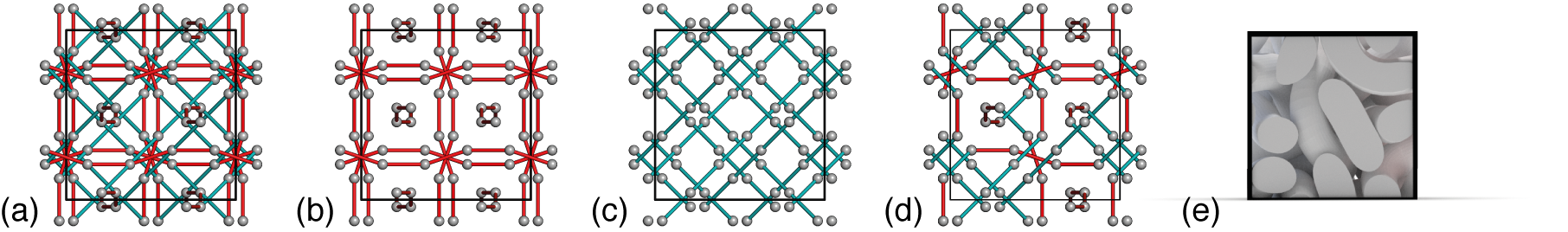}
\end{center}
{\bf Figure 2. Generation of disordered entanglements.} (a) Representation of the parent $Ia\bar3d$ network from which our entanglements are derived. Nodes are shown as grey spheres and occupy the $48g$ Wyckoff positions. These nodes are connected to one another by a set of two symmetry-distinct types of linkers, shown here in red and teal. Two red linkers and two teal linkers meet at each node (see SI). Entanglements are generated by partitioning this linker set into two subsets of equal size. The trivial partitioning gives the (b) $\Pi^\ast$ and (c) $\Sigma^\ast$ topologies, corresponding to the cases of all-red and all-teal linkers, respectively. Partitionings involving a combination of red and teal linkers are predominantly disordered, such as the example shown in (d). Subsequent geometric relaxation as described in the text and SI yields a disordered entanglement (e).

\clearpage
\begin{center}
\includegraphics{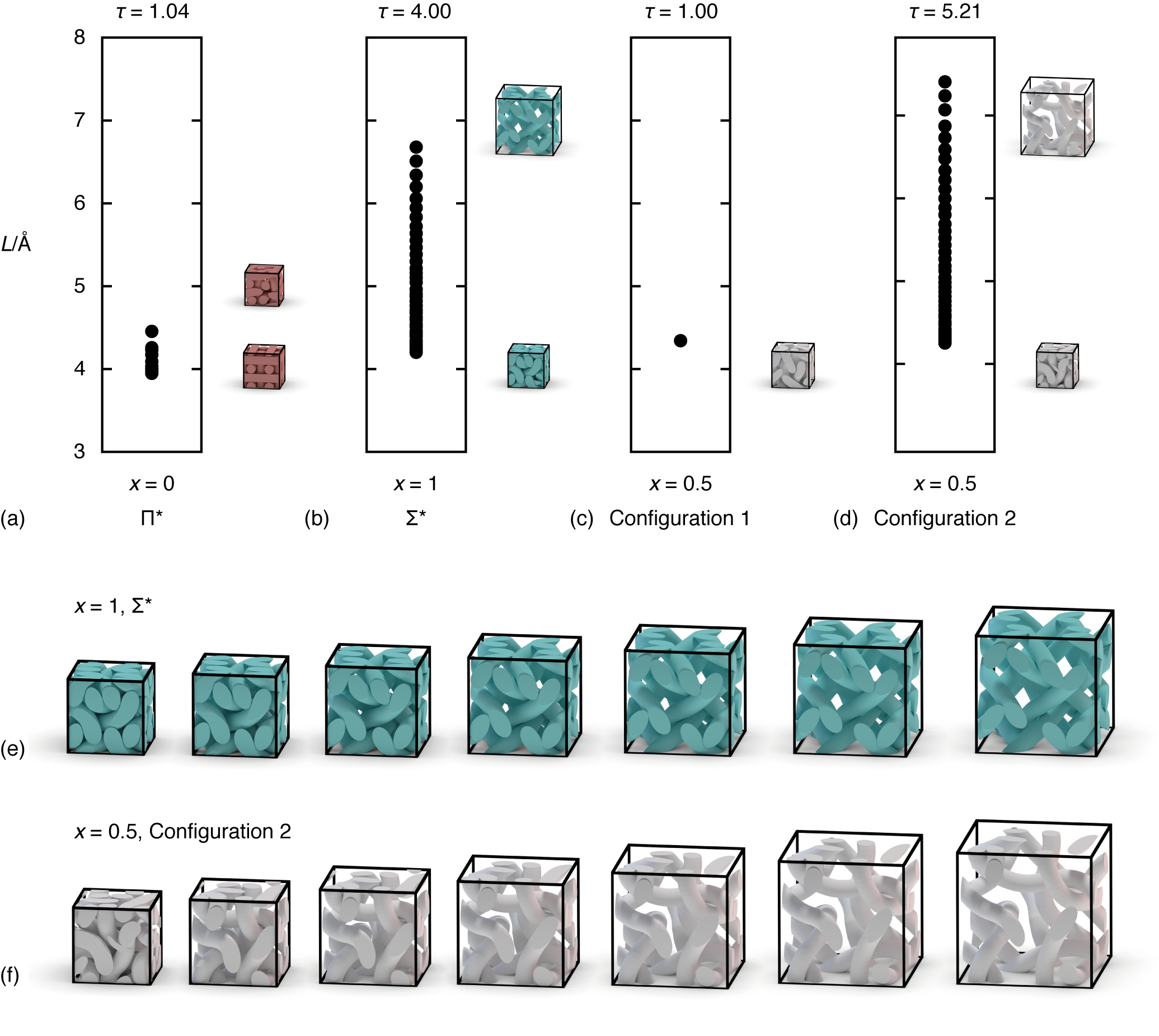}
\end{center}
\clearpage
{\bf Figure 3. Swelling of ordered and disordered entanglements.} Panels (a)--(d) show the range of relative unit cell dimensions accessible to entanglements with different topologies: (a) $\Pi^\ast$, (b) $\Sigma^\ast$, (c) a disordered entanglement with $x=0.5$ and minimal swelling capacity, and (d) a disordered entanglement also with $x=0.5$ but with an extreme swelling capacity. Representations of the collapsed and extended entanglements are shown on the right of each plot. (e) Illustration of the swelling mechanism for the $\Sigma^\ast$ weaving, as described in Ref.~\citenum{Evans_2011}. (f) The corresponding mechanism for the disordered entanglement of panel (d). Note that in both (e) and (f) fibres remain in contact with one another, but pore volume increases as the entanglement curvature is decreased.

\clearpage
\clearpage
\begin{center}
\includegraphics{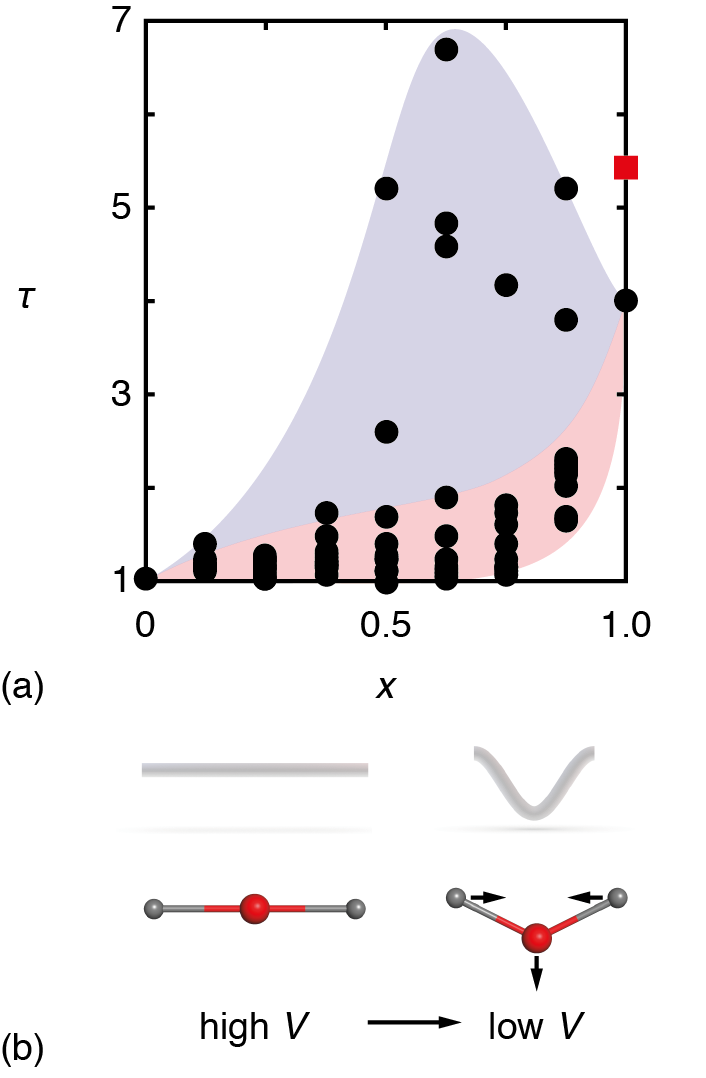}
\end{center}
{\bf Figure 4. Disorder/swelling relationship.} (a) Variation in swelling capacity $\tau$ with degree of topological disorder $x$. The value $\tau(1)$ is given both for the empirical calculation of our study (black circle) and for the DFT result of Ref.~\citenum{Evans_2015} (red square). Our data show that disorder mostly quenches swelling capacity (shaded red area) but that certain disordered topologies can show extreme capacities (shaded blue area). (b) Correspondence between variation in fibre curvature and the tension mechanism of negative thermal expansion (NTE). In NTE materials, increased transverse displacements at high temperatures act to reduce volume by increasing the effective curvature of the framework backbone.

\end{document}